\documentclass[apj]{emulateapj}
\usepackage{apjfonts}
\usepackage{epsfig}

\lefthead{Paladini et al.}
\righthead{The luminosity function of Galactic HII regions}

\begin{document}

\title{Re-analysis of the radio luminosity function of Galactic HII regions}

\author{
R. Paladini\altaffilmark{1,*},
G. DeZotti\altaffilmark{2,3},
A. Noriega-Crespo\altaffilmark{1},
S.J. Carey\altaffilmark{1}
}

\altaffiltext{*}{Corresponding author: paladini@ipac.caltech.edu}
\altaffiltext{1}{Spitzer Science Center, 1200 E. California Blvd., Pasadena, CA. 91125 - USA}
\altaffiltext{2}{INAF - Osservatorio Astronomico di Padova, Vicolo dell'Osservatorio 5, I-35122 Padova, Italy}
\altaffiltext{3}{SISSA, Via Beirut 4, I-24014 Trieste, Italy}

\begin{abstract}

We have re-analyzed continuum and recombination lines radio data available in the literature in order to derive the luminosity 
function (LF) of Galactic HII regions. The study is performed by considering the first and fourth Galactic quadrants independently. We estimate the completeness 
level of the sample in the fourth quadrant at 5 Jy, and the one in the first quadrant at 2 Jy. We show that the two 
samples (fourth or first quadrant) include, as well as giant and super-giant HII regions, a significant number of sub-giant sources. 
The LF is obtained, in each Galactic quadrant, with a generalized Schmidt's estimator using an effective volume derived from the 
observed spatial distribution of the considered HII regions. The re-analysis also takes advantage of recently published ancillary absorption data allowing to 
solve the distance
ambiguity for several objects. A single power-law fit to the LFs retrieves a slope equal to -2.23$\pm$0.07 (fourth quadrant) and 
to -1.85$\pm$0.11 (first 
quadrant). We also find marginal evidence of a luminosity break at $L_{knee}$ = 10$^{23.45}$ erg s$^{-1}$ Hz$^{-1}$ for the LF in 
the fourth quadrant. 
We convert radio luminosities into equivalent H${\alpha}$ and Lyman continuum luminosities to facilitate comparisons with extra-galactic 
studies. We obtain an average total HII regions Lyman continuum luminosity of 0.89 $\pm 0.23 \times 10^{53}\,\hbox{sec}^{-1}$, corresponding 
to 30$\%$ of the total ionizing luminosity of the Galaxy.

\end{abstract}


\section{Introduction}

Extensive studies have investigated the LF of HII regions in external galaxies. These, mostly based
on H$\alpha$ observations, have established that: 1) the slope of the LF is  
correlated with Hubble type, with early-type galaxies having a steeper slope 
than late-type (Kennicutt, Edgar $\&$
Hodge 1989, hereafter KEH89);  2) the LF becomes significantly flatter at low 
luminosities (KEH89; Rand 1992; Walterbos $\&$ Braun 1992; Rozas et al. 1996, 
1999, 2000; Bradley et al. 2006). Such observational facts have been
interpreted in terms of evolutionary effects (Oey $\&$ Clarke 1998). Variations
of the LF in spiral arm versus inter-arm regions are controversial: some
studies claim that the LF of interarm HII regions is steeper (KEH89; Banfi et
al. 1993; Rand 1992), but others do not support this conclusion (Rozas et al.
1996; Knapen et al. 1993, Knapen 1998).

Remarkably, there is a surprising paucity of similar studies regarding the HII regions 
population of our own Galaxy. This is partly due to the fact that HII regions lie on the 
Plane, where H$\alpha$ is heavily obscured by dust, therefore limiting the observation 
to HII regions located within a radius of a $\sim$ kpc. Radio surveys could in principle 
circumvent this problem but, in the past, only a few have uniformly targeted the Galactic Plane region 
with sufficient angular resolution and sensitivity to resolve individual sources. Likely the near future 
will see a breakthrough in this direction, with the completion of extensive radio surveys carried out 
through interferometric techniques, such as the International Galactic Plane Survey (IGPS) at 1.4GHz and 408 MHz, or 
MAGPIS (White et al. 2005), at 5 and 1.4 GHz. Kerton et al. (2007) have published a catalog of Galactic extended sources 
from the Canadian Galactic Plane survey (Taylor et al. 2003), part of IGPS. 
However, despite a remarkable work of detection and extraction, the sources have not been classified yet.

The reference  works for the LF of Galactic HII regions are Smith $\&$ Kennicutt (1989) 
and McKee $\&$ Williams (1997). They both rely on the list of HII regions
compiled by Smith, Biermann $\&$ Mezger (1978, hereafter SBM). This list is a compilation of 
93 HII regions and it is characterized by important caveats. For example: the Southern sources in the list (49 
out of 93) are mainly from the 5 GHz recombination lines survey  by Wilson et al. (1970). These authors claim that 
their survey is complete for sources having a peak antenna temperature of 1.3 K, or greater, at 5 GHz. However, this statement 
is based on the continuum data which have been obtained through a variety surveys, carried out with different frequencies
and angular resolutions. For instance, the Thomas $\&$ Day (1969a,b) 2.7-GHz survey
has a spatial resolution of 8.2$^{\prime}$, while the Hill (1968) 1.4-GHz survey is characterized by 
a resolving power of 14$^{\prime}$. Most importantly, the completeness levels of these continuum surveys appears to be
unknown. 

Therefore, Wilson et al. cannot be certain that {\em{all}} sources above the 1-K thereshold have been
picked up, in the first place, by those surveys. 
This means that the recombination lines survey has an intrinsic bias, with important consequences 
on the completeness for giant and super-giant HII regions, as we will discuss in \S~5. 

Complementary studies have focused on the LF of {\em ultracompact} HII regions (Comeron $\&$ Torra 1996; Casassus et al. 2000). 
The work by Comeron $\&$ Torra is probably the only attempt found in the literature
to take into account the complex spiral-arm geometry of our Galaxy. However, since it relies   
only on the IRAS point-source catalog (PSC) and lacks any kinematic information on the sources,
it purely consists in a multi-parameter fit analysis, which has obvious limitations.

For the reasons described above, we have decided to look in the literature and 
search for alternative data to the SBM list in order to obtain an independent 
derivation of the LF of Galactic HII regions. We are 
in fact convinced that archival data are widely and unjustifiably under-used in this field and, until 
new catalogs of HII regions are made available, for instance, by the MAGPIS or the IGPS consortia, these archival data should be fully exploited.  This is 
even more true given the new release of ancillary data, such as of those on HI absorption, which 
make it possibile, for many of these sources, to break, for the first time, the 
distance degeneracy. 

In the following, we will consider two definitions of the LF denoted,
respectively, by $n(L)\,dL$ and  $\phi(L)\, d\log L$. The former gives the
number of sources per unit volume with luminosity in the range [$L$, $L+dL$]
and is usually parameterized as $n(L)\, dL \propto L^{-\alpha}\, dL$; the
latter gives the number of sources per unit volume with luminosity in the
logarithmic interval [$\log L$, $\log L + d \log L$] and is represented as
$\phi(L)\ d \log L \propto L^{-\alpha+1}\, d\log L$.

\section{Sample selection}

The vast majority of catalogued HII regions is located in the first ($0^\circ <
l < 90^\circ$) and fourth ($270^\circ < l < 360^\circ$) Galactic quadrants. 

The most recent, extensive radio recombination lines surveys in the Galactic Plane are the Caswell $\&$     
Haynes (1987, hereafter CH87) survey in the fourth quadrant, and  the 
Lockman (1989) survey in the first quadrant. 

The CH87 survey has covered the region $210^\circ < l <
360^\circ$, $|b| < 2^\circ$, in the fourth quadrant. With an angular resolution
of $\sim 4.1^\prime$, it represents the follow-up of the continuum 5 GHz survey by
Haynes et al. (1978, 1979), which has been carried out at the same angular resolution as the recombination 
lines survey.  The CH87 survey consists of H109$\alpha$ and H110$\alpha$     
recombination lines measurements of {\em most} of the sources (as stated by the same authors) with peak brightness temperature    
exceeding 1 K at 5 GHz, equivalent to a completeness 
limit, in peak flux density, of  1.3 Jy. 
CH87 provide, for each  HII region in their surveyed region, the measured recombination 
line velocity and electron temperature, as well as the angular diameter
and the integrated flux density as observed in the continuum 5-GHz survey. We point out that these 
flux densities and diameters have been derived by Haynes et al. by measuring by hand contours of 
surface brightness. The same method has been applied to the sources in the SBM list located in the Southern hemisphere, as 
described in Wilson et al. (1970).

The Lockman (1989) survey, with a spatial resolution of 3$^\prime$, is a follow-up of the 
4.8 GHz Altenhoff et al. (1979) continuum survey, made with comparable angular resolution (2.6$^\prime$).  
It consists of observations of the
H87$\alpha$, H88$\alpha$ and H85$\alpha$ hydrogen recombination lines, and it 
is complete down to 1 Jy, in peak flux density, for $0^\circ < l < 60^\circ$, while 
for $ l > 60^\circ$, the completeness of
the survey is less uniform. Interestingly, the Lockman survey has covered, with equal 
angular resolution, the same region 
of the sky of the H110$\alpha$ and formaldehyde (H$_{2}$CO) absorption line survey by Downes et al. (1980, hereafter DWBW). 
DWBW also claim to be complete to a peak flux density of 1 Jy, although Lockman reports
that $\sim 200$ sources are observed for the first time. This fact could be
explained with misidentification problems in the DWBW survey. 
For the Lockman sources, flux densities and angular diameters are taken from the 4.8 GHz Altenhoff et al. (1979) survey, and 
these have been measured through automated gaussian fits. The same technique has been used to derive flux and angular 
sizes for the sources in the SBM compilations located in the Northern hemisphere, as discussed in Reifenstein 
et al. (1970). 

Instrumental systematics can potentially degrade in a significant way 
continuum measurements, especially with the technology available 20 or 30 years ago. One of the 
main problems is represented by $1/f$ noise which can cause fluctuations of the baselines and make the measured 
flux densities unreliable. The solution is typically to scan the sky at a fast rate. The 
scan rates of the Haynes et al.'s and Lockman's surveys are 2.5$^\circ$/min and 80$^{\prime}$/min, respectively. 
In comparison, the sources above 1 K at 5 GHz and selected by Wilson et al. for their recombination lines survey,
 have been re-observed in the continuum 
by Goss $\&$ Shaver (1970) at a rate of  1$^\circ$/min, meaning that 
the uncertainties in the flux densities quoted by CH87 and Altenhoff et al. (1979), due 
to effects such as $1/f$ noise, are lower or at least of the same order as the values 
reported by SBM. 

We note that Paladini et al. (2003; hereafter Paper I), compiled a  radio catalog of 1442 
Galactic HII regions for which angular diameters and flux densities
at a reference frequency, namely 2.7 GHz, are given. This is the so-called
Master Catalog, in which, for each source,  the quoted values at the reference
frequency are obtained by taking the weighted mean over all the available data
reported in the literature. This procedure also requires to estimate
measurement errors when these are not provided by the authors of the original
surveys. For $\sim$ 800 of the HII regions in the catalog, radio recombination
lines have been measured and the mean line velocities, line and electron
temperatures are reported in the Master Catalog. Such a data set, combined with the Fich, Blitz $\&$ Stark (1989,
hereafter FBS89) rotation curve, has been used in Paladini, Davies $\&$ De
Zotti (2004; hereafter Paper II) to compute, for a subset of 575
sources{\footnote{Only the sources with a measured velocity $|V_{lsr}| >$ 10
km/s are considered in the analysis.}}, galactocentric and solar distances. 

We could have in principle used the Paladini et al.'s catalog to perform 
our analysis of the LF. However, as mentioned above, the Master Catalog contains {\em weighted averages} of
measurements obtained with different instruments, at various frequencies and
angular resolutions. This fact {\em per se} can introduce complications in, for 
example, assessing the actual completeness of the sample. 

On the contrary, the CH87 and Lockman's 
data bases present the advantage (over the Paladini et al.'s catalog but also over the SBM list) 
of being {\em uniform} samples, obtained by observations of large areas of the sky with a single 
instrument, characterized by a given sensitivity and angular resolution.

\section{Completeness of the samples}

CH87 observed 316  HII regions in total. The authors state in the paper that they have likely observed most of the 
sources with $S_{p} >  1.3\,$Jy. The first step is then to remove from the sample the sources below this threshold, i.e. 
since CH87 quote, for each source, the  
integrated flux at 5 GHz, $S_{i}$, this corresponds to exclude the sources for
which:

\begin{equation}
S_p = S_i \frac{{\theta_b}^2}{{\theta_b}^2 + \theta_{\rm HII}^2} \le 1.3 \hskip 0.2truecm \hbox{Jy}
\end{equation}

where $\theta_{\rm HII}$ is the angular diameter of the source, and $\theta_{b}$ is the angular resolution of the survey (4.4$^{\prime}$).   
The application of the above criterium leaves us with 238  HII regions. At this point, we still need to establish 
the actual completeness of the survey in terms of flux density. For this purpose, we consider the 
observed differential counts (see Fig.~1) of the sample and we fit to this, along the lines of Paper II, a simple function of the type $S^{-1}$, representing the scaling of the number of 
sources, as 
seen from an observer located at the Sun, if these were distributed along a disk. This approximation is valid within the range 30 Jy $> S_{5GHz} >$ 5 Jy. For 
$S_{5GHz} >$ 30 Jy, the distribution steepens and is better represented by a function $S^{-3/2}$, appropriate for a uniform three-dimensional 
distribution, roughly corresponding to nearby sources. Errors have been computed, in each flux bin, by applying the formulae by Gehrels (1986).
Specifically, the upper limits are given by:
\begin{equation}
\lambda_{u} \simeq n + S \sqrt{n + \frac{3}{4}} +  \frac{S^{2} + 3}{4}
\end{equation}
while the lower limits are:
\begin{eqnarray}
\lambda_{l} \simeq n{\left(1 - \frac{1}{9n} - \frac{S}{3\sqrt{n}} + \beta n^{\gamma}\right)}^{3}
\end{eqnarray}
where $S =1$, $\beta =0$ and $n$ is the number of sources per bin. By removing 
from the sample the objects with an integrated flux $<$ 5 Jy, we are left with 162 sources. 
In the following, we will derive the LF of the fourth quadrant also for flux cut-offs different (and above) the cut-off value obtained from the the analysis of the 
differential counts. These additional LFs will be used for comparison with the one obtained for the reference 5-Jy cut-off, in order to estimate the potential 
error introduced by the assesment of the completeness level for the CH87 sample.

\begin{figure}[h]
\plotone{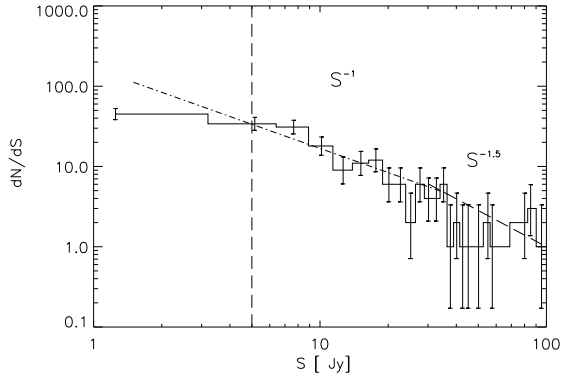}
\epsscale{.6}
\caption{Differential source counts for the CH87 sample. The fits for a disk (dashed-dotted line) and spherical (dashed) 
distributions are shown. The completeness level is 
denoted by the dashed vertical line.}
\end{figure}

The Lockman (1989) sample has 256 sources within longitude 2$^\circ$ $< l < $ 60$^\circ$. After removing the sources 
with peak flux density $S_{p} <  1\,$Jy, or for which:

\begin{equation}
S_p = S_i \times \frac{{\theta_b}^2}{{\theta_b}^2 + \theta_{\rm HII}^2} \le 1\,\hbox{Jy}
\end{equation}

(taking $\theta_b = 2.6^{\prime}$), we are left with 115 sources. In this case, the source counts are well fitted by a model 
$\sim$ $S^{-1}$ in the flux range 30 Jy $> S_{5GHz}$ $\simeq$ 2 Jy, while for  $S_{5GHz} >$ 30 Jy a function $S^{-3/2}$ is a better representation of the data. 
The exclusion of sources below the 2-Jy thereshold leaves 93 objects. Along the same lines of the CH87 sample, we will also compute the LF in the first 
quadrant for completeness cut-offs above the one estimated from the differential counts.

\begin{figure}[h]
\plotone{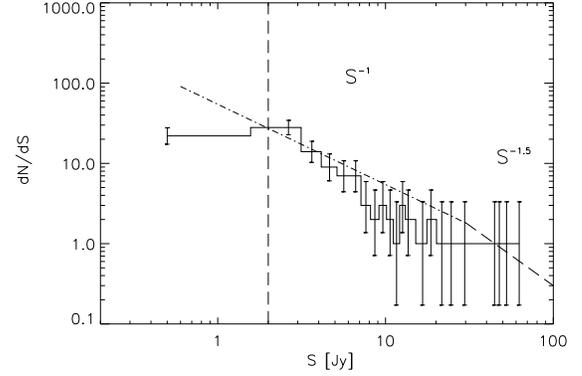} 
\epsscale{0.7}
\caption{Differential source counts for the Lockman (1989) sample. The fits for a disk (dashed-dotted line) and spherical (dashed)  
distributions are shown. The completeness level is denoted by the dashed vertical line.}                                                                  
\end{figure}

\section{Exclusion of sources with $|V_{LSR}| <$ 10 km/s}

Sources with radial velocities $|V_{\rm LSR}| > 10$ km/s can, in principle, be affected by non-circular
streaming motions, but, as noted for instance by Quireza et al. (2006a), an
accurate quantification of this effect is not possible given our current
knowledge of the Galactic velocity field. For this reason, throughout the text,
we have assumed that the circular motion hypothesis is generally valid.

On the contrary, sources with measured radial velocities 
$< 10$ km/s are known to be significantly contaminated by
peculiar velocity components, and distances derived by using  these
velocities can be highly unreliable.
 
Therefore, from each sample (CH87 and Lockman (1989)), we remove the sources with $|V_{\rm LSR}| < 10$ km/s. For the
reference cases of flux cut-offs of 5 and 2 Jy, this corresponds to remove 23 objects in the fourth
quadrant and 5 objects in the first quadrant. 

In \S~8.3, we will examine the effect, on the derived LF, due to the removal of the low-velocity sources.

\section{Radial and solar distances}

For each source in the CH87 and Lockman (1989) samples, we compute the galactocentric and solar distances, R and D, respectively,
by combining the measured hydrogen recombination line velocity with the rotation curve by FBS89. 
While in Paper II we used $R_{0} = 8.5\,$kpc, we have adopted here a somewhat smaller value, $R_{0} = 8\,$kpc,
favoured by recent determinations (Eisenhauer et al. 2003, 2005).
FBS89 provide a useful tabulation of the rotation curve for
several values of $R_{0}$ (see Table~4 of their paper), including $R_{0} = 8.0\,$kpc. In \S~8.2, we will show that
the adoption of rotation curve models other than FBS89 has a very minor impact on the derived LF.

As for the solar distances, part of the HII regions in both samples are
affected by the well-known distance degeneracy problem. These are the sources within the solar circle
for which two solutions are consistent with the observed radial velocity. In this case, we proceed
as in Paper II, using auxiliary (optical or absorption) data, when available, to break the degeneracy. 

For the fourth quadrant (CH87 sample), with respect to
Paper II, we have complemented the list of references of auxiliary data with 
the work on HI absorption by Fish et al. (2003), Nord et al. (2006) and Quireza et al. (2006b). 

When sources lack auxiliary data, we make use of the luminosity-physical diameter
correlation found in Paper II and, in particular, of eq.~(7) in that paper, which, for the 5 GHz case takes the form:

\begin{equation}\label{eq:LD}
D = C \times \hskip 0.1truecm 10^{a\over 2-b} \left(\theta\over {1^{'}}\right)^{b\over 2-b}\left(\nu\over{1\, 
\hbox{Hz}}\right)^{-1\over{2-b}}\left(S\over{1\,\hbox{Jy}}\right)^{-1\over{2-b}} \!\!\!\! \hbox{kpc}
\end{equation}

\noindent with $C = 1.2767 \times 10^{-19}$, $a=32.4$ and $b=0.86$.{\footnote{Note that, in Paladini et al. (2004), eq.~(7) is missing the 
factor $1.2767 \times 10^{-19}$ and 
the frequency $\nu$ is in wrong units, i.e. GHz instead of Hz.}}  
For the reference case of a flux cut-off at 5 Jy, we have applied eq.~(5) to 61 out of 139 sources, corresponding 
to 44$\%$ of the total. 

We emphasize that the relation above is
used to choose between the near and far kinematic solutions for a given
measured line velocity, not to compute a heliocentric distance from the
observed angular diameter and flux. Following Paper II, we performed on the
CH87 sample of sources a test to cross-check whether the application of eq.~(2)
to the sources with auxiliary data assigns the correct solution. The test shows
that the luminosity-diameter correlation allows to choose the {\em correct}
solution in $\sim$ 80$\%$ of the cases.

\begin{figure}[h]
\plotone{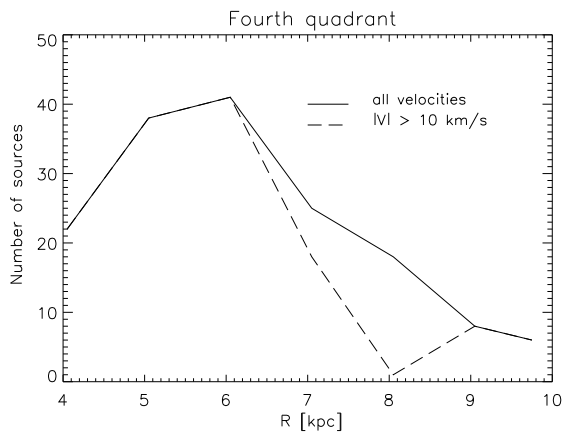}
\epsscale{.6}
\caption{Radial distribution of HII regions in the 4th Galactic quadrant. Sources 
are from CH87, and their flux density is above 5 Jy at 5 GHz. 
Bins of $\Delta R$ = 1 kpc. The solid line denotes the distribution obtained when sources 
with $|V_{LSR}| <$ 10 km/s are left in the sample, while the dashed line corresponds to the case when 
these sources are removed.}
\end{figure}

In the first quadrant, 79 out of 88 of the Lockman sources above the reference flux cut-off at 2 Jy lie inside the solar circle. 
The first step is to make use of auxiliary 
(optical and absorption) data to solve the kinematic distance ambiguity. As for
the CH87, we complement the list of references of Paper II with 
the HI absorption data by Fish et al. (2003),
Kolpak et al. (2003), Quireza et al. (2006a), Anderson $\&$ Bania (2009), as
well as with the formaldehyde (H$_{2}$CO) absorption data provided by Araya et
al. (2002), Watson et al. (2003), Sewilo et al. (2004) and the same DWBW.
However, as noted by Anderson $\&$ Bania, HI absorption methods are usually
more robust than H$_{2}$CO to break the  distance degeneracy, given that HI is
more ubiquitous than H$_{2}$CO. For this reason, whenever a source has both
H$_{2}$CO and HI absorption data, we favor the solution associated to HI.
Following these criteria, we are able to solve the distance ambiguity for 70 
objects. To the remaining 9 (10$\%$ of the total) sources we have applied
eq.~(5). 

The radial distributions of the HII regions above the flux cut-offs of 5 and 2 Jy 
are illustrated in Fig.~3 and Fig.~4, for the fourth and first quadrant, respectively. 
Note that the figures show the radial distribution both 
when sources with $|V_{LSR}| <$ 10 km/s are removed or left in the samples. Clearly, 
the removal of these sources causes, in the fourth quadrant, an artificial dip in the distribution for $R \sim$ 8 kpc.

Remarkably, and regardless of the exclusion of the low-velocity sources, the radial distribution 
appears to be different in the two quadrants. As emphasized by several authors (e.g. Lockman 1981,  
1989; CH87), this is a consequence of the fact that HII regions, as
massive stars in general, trace the spiral structure of the Galaxy and, as
such, reflect in their spatial distribution the asymmetry of the arms geometry
as seen by an observer located at the Sun position. 

\begin{figure}[h]
\centering
\plotone{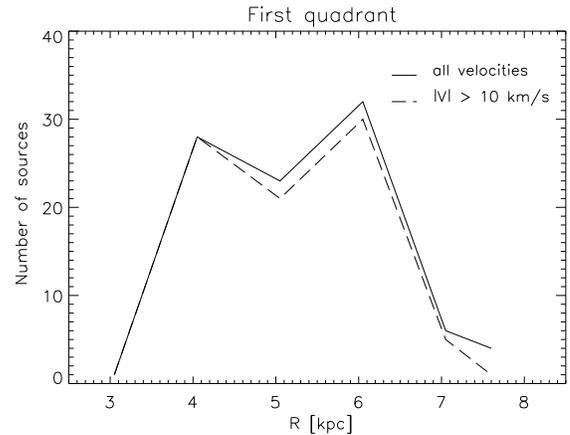}
\epsscale{.6}
\caption{
Radial distribution of HII regions in the 1st Galactic quadrant. Sources
are from Lockman (1989), and their flux density is above 2 Jy at 4.8 GHz.
Bins of $\Delta R$ = 1 kpc. As for the fourth quadrant: the solid line denotes the distribution obtained 
by including all sources; the dashed line denotes the one corresponding to the removal of sources 
with $|V_{LSR}| <$ 10 km/s.}
\end{figure}

\section{Composition of the samples}

The SBM compilation used in Smith $\&$ Kennicutt (1989) and McKee $\&$ Williams (1997) is supposedly complete, over the whole Galaxy, for 
super-giant HII regions, and for our side of the Galaxy, for giant HII regions. SBM define super-giant HII regions those having a 
Lyman continuum luminosity $N_{c} >$ 50$\times$ 10$^{49}$ s$^{-1}$, while giant HII regions are 
the ones for which 4$\times$ 10$^{49}$ s$^{-1} > N_{c} >$ 50$\times$ 10$^{49}$ s$^{-1}$. These definitions are provided taking as a 
reference the luminosity of Orion for $R_{0} = 10\,$kpc and $T_{e} = 10000\,$K. Since in our 
derivation of the radial and solar distances we use $R_{0} = 8.\,$kpc and $T_{e}$ $\sim 7000\,$K, 
we rescale, following McKee $\&$ Williams, the Lyman 
luminosities by a factor 0.75 which takes into account both corrections. Accordingly, the new limits for super-giant and giant HII regions become, respectively, 
$N_{c} >$ 37.5$\times$ 10$^{49}$ s$^{-1}$ and 3$\times$ 10$^{49}$ s$^{-1} > N_{c} >$ 37.5$\times$ 10$^{49}$ s$^{-1}$.  In addition, 
sub-giant HII regions will be the sources with an ionizing luminositiy less than 3$\times$ 10$^{49}$ s$^{-1}$. 

We use the flux densities tabulated by Haynes et al. (1978) for the CH87 sources and those by Altenhoff (1978) for the 
Lockman sample, combined with the solar distances assigned as described in the previous sections, to compute for each 
HII region in the fourth and first quadrant its ionizing luminosity, following the Condon's (1992) relation: 

\begin{eqnarray}
Q = 6.3 \times 10^{52} \hbox{photons}\,\hbox{s}^{-1} \left(T_{e}\over{10^{4} K}\right)^{-0.45} \cdot \\ \nonumber
 \cdot \left({\nu\over{\rm GHz}}\right)^{0.1} {L_{\nu}\over 10^{27}\,\hbox{erg}\,\hbox{s}^{-1}\,\hbox{Hz}^{-1} } \nonumber
\end{eqnarray}
%
where, for the electron temperature, we use $T_{e}$ = 6200 K for the sources in the fourth quadrant, and $T_{e}$ = 7400 K 
for the objects in the first quadrant. These numbers represent mean values of the electron temperatures of the  sources 
in the CH87 and Lockman's samples. 

Applying the definition of sub-giant, giant and super-giant HII regions given above, we obtain the composition of the 
CH87 and Lockman samples provided in Table~1.

\begin{deluxetable}{ccc}
\tabletypesize{\footnotesize}
\tablewidth{0pt}
\tablecaption{Composition of CH87 and Lockman samples}
\tablehead{ 
\colhead{Type} &
\colhead{CH87} &
\colhead{Lockman (1989)}\\
}
\startdata
sub-giant HII regions & 74 &  51\\
giant HII regions & 62 &  35  \\
super-giant HII regions & 3 & 2 \\
\\
\enddata
\tablecomments{Quoted values are for the reference cases of a flux cut-off at 5 and 2 Jy, for the CH87 and Lockman (1989) samples, 
respectively.}
\end{deluxetable}

Clearly, in both quadrants we have a significant contribution from sub-giant HII regions. This is important given that, 
as noted by McKee $\&$ Williams (1997), most of the ionizing luminosity of the Galaxy is likely contributed 
by small HII regions, rather than by the super-giant and giant populations. 
Furthermore, we note that, for the fourth quadrant, the SBM list quotes 49 between giant and super-giant HII regions, while 
the CH87 sample includes 65 in total. Likewise, in the first quadrant, the SBM compilation has 33 giant and super-giant HII regions 
in the region $2^\circ < l < 60^\circ$, compared to 37 for the Lockman (1989) sample. This shows that the samples 
considered here are likely more {\em complete} than the SBM catalog, as mentioned in the Introduction.

\section{The LF for an inhomogeneous source distribution}

One of the limitations of previous works which have investigated the LF
for Galactic HII regions is that they did not allow for a realistic geometric distribution of the sources.
In this section we describe the approach adopted for our analysis. Under the assumption that the LF
is independent of position, the number $N(L)$ of HII regions with luminosity $L$ within $d\log L$
detected above a flux threshold $S$ is related to the LF, $\phi(L)\,d\log L$, by:
\begin{eqnarray}\label{eq:NL}
N(L)\,d\log L = \phi(L)\, d\log L \int^{D_{\rm max}(S|L)}_{0} {dD\, D} \\
          \int^{l_{\rm max}}_{l_{\rm min}}{dl\, n(R)} \int^{z_{\rm max}}_{z_{\rm min}} {dz\, \rho(z)} \nonumber
\end{eqnarray}
with $R = (R_{0}^2 + D^2 - 2 D R_{0}\cos l)^{1/2}$. In the above expression $n(R)$ and $\rho(z)$
describe the radial and vertical density profiles, respectively, and the multiple integral
on the right-hand side is the {\em effective volume} within which a source of luminosity $L$ is included in the catalog, i.e.:
\begin{eqnarray}\label{eq:Veff}
V_{\rm eff}(L|S)=\int^{D_{\rm max}(S|L)}_{0} {dD\, D} \\
             \int^{l_{\rm max}}_{l_{\rm min}}{dl\, n(R)} \int^{z_{\rm max}}_{z_{\rm min}} {dz\, \rho(z)} \nonumber
\end{eqnarray}
where $D_{\rm max}(S|L) = (L/ 4\pi S)^{1/2}$, $z_{\rm max} = D \sin(b_{\rm max}) \simeq D\, b_{\rm max}$
and $z_{\rm min} = D\, \sin(b_{\rm min}) \simeq D\, b_{\rm min}$.

From eqs.~(7) and (8) we can define a generalized Schmidt's (1968) estimator for the
LF:
\begin{equation}
 \phi(L) \Delta\, \log L = \sum_{\rm i} {1\over{V_{\rm eff,i}(L|S)}}
\end{equation}
where the sum is over all sources with luminosity within the interval $\Delta \log L$ around $L$.
In eq.~(8), for the vertical density profile we use a function of the form:
\begin{equation}
\rho(z|R) = \exp[-0.5(z/\sigma_z)^2]
\end{equation}
If $z_{\rm max}$ is much larger than the vertical scale length $\sigma_{z}$ and
$z_{\rm min} = -z_{\rm max}$, the integral over $dz$ in the right-hand side of
eq.~(8) gives $(2\pi)^{1/2}\sigma_{z}$. 

Bronfman et al. (2000) have investigated the scale height distribution of a population 
of 748 OB stars, for which (their Table~2) they tabulate values at different Galactocentric 
radii, for the North and South hemispheres, both separately and combined. We have fitted their 
quoted values for the combined distribution, after correcting for $R_{0}$ = 8 kpc. 
The data result to be well represented by a function of the form:

\begin{equation}
z_{1/2} = 2.85 R^{1.69}\,\hbox{pc},
\end{equation}

\noindent
where $z_{1/2}=(2\ln2)^{1/2}\sigma_z$. This relation is also consistent with 
the range of values (39 to 52 pc) quoted in Paper II for HII regions in our Galaxy, therefore  
we adopt it.

The choice of the radial density profile is more complex (see Fig.~12 of Paper II). 
In \S~5 we have derived the distribution 
in Galactocentric distances of the sources in our samples located in the fourth (Fig.~3) and first (Fig.~4) quadrants. 
Based on these distributions, we model the $n(R)$ as: 

\begin{eqnarray}\label{eq:2Gaus}
n(R) = \exp\left[-((R-R_{pk1})/\sigma_{r1})^2\right] + \\
       + A\exp\left[-((R-R_{pk2})/\sigma_{r2})^2\right] \nonumber
\end{eqnarray}

In the expressions above, R$_{pk1}$, R$_{pk2}$, $\sigma_{r1}$, $\sigma_{r2}$ and A are free parameters to be determined
fitting the observed distribution of sources as a function of the galactocentric distance $R$, given by:
\begin{eqnarray}\label{eq:NR}
{\cal N}(R)\,\Delta R=n(R)\,R\,\Delta R \int_{\theta_{\rm min}(R)}^{\theta{\rm max}(R)}d\theta\, \cdot \\
\cdot \int_{z_{\rm min}(R,\theta)}^{z_{\rm max}(R,\theta)}dz\,\rho(z,R) \int_{\log L_{\rm min}(R,\theta)}^{\log L_{\rm max}}d\log L\,\phi(L)\ \nonumber
\end{eqnarray}

We note that the adoption of a double-Gaussian profile for the first quadrant is motivated by the   
fact that the line of sight intercepts the Carina  and Scutum-Crux spiral arms. 

To keep the number of parameters within a manageable limit, we adopt, in
eq.~(13), the usual power-law representation for the luminosity function
$\phi(L)\,d\log L \propto L^{-1}\, d\log L$ (i.e. $\alpha = 2$), after having
checked that the results are little affected by changes of its slope within the
currently accepted range, $1.5 \le \alpha \le 2.5$. The estimate of $\phi(L)$
is thus based on an iterative approach: we use a first guess on the shape of
$\phi(L)$ to derive the parameters for $n(R)$ that we then exploit to obtain a
better estimate for the LF. More details on the calculation are provided in the
Appendix.

The minimum $\chi^{2}$  values of the parameters for both samples are listed in Table~2 and 3. 
In view of a derivation of a LF for different 
levels of completeness, we have derived the best-fit parameters for the radial distribution 
for each value of flux cut-off. 
For the reference cases of flux cut-offs at 5 and 2 Jy for the CH87 and Lockman's samples, we obtain that 
the best fits correspond to a $\chi_{r}^{2}$  
of 2.08 and 1.13, for a bin size  $\Delta$R = 1 kpc.  
The value of $\chi_{r}^{2}$ is actually an
overestimate and does not denote an unacceptably poor fit because the adopted
errors correspond only to the Poisson fluctuations of the number of sources in
each bin and do not include the contributions from the uncertainties in the
estimates of distances $R$. Thus, although we are well aware that the axially
symmetric distributions of eq.~(12) are certainly an oversimplification,
a more refined model, entailing more parameters, does not appear to be
warranted in the present data situation.

\begin{deluxetable}{c|ccc}
\tabletypesize{\footnotesize}
\tablewidth{0pt}
\tablecaption{Best-fit parameters of the radial profile for the 4th quadrant}
\tablehead{
\colhead{Parameter} & \colhead{5 Jy} & \colhead{7 Jy} & \colhead{9 Jy}\\
}
\startdata
$R_{pk1}$ & 4.28  (kpc) & 4.26 (kpc) & 4.27 (kpc) \\
$R_{pk2}$ & 9.54 (kpc) & 9.53 (kpc) & 9.53 (kpc) \\
$\sigma_{r1}$ & 1.44 (kpc) & 1.45  (kpc) & 1.45 (kpc) \\
$\sigma_{r2}$ &  0.42 (kpc) & 0.42 (kpc) &  0.42 (kpc)    \\
A & 0.30  & 0.29 & 0.30 \\
\enddata
\end{deluxetable}

\begin{deluxetable}{c|ccc}
\tabletypesize{\footnotesize}
\tablewidth{0pt}
\tablecaption{Best-fit parameters of the radial profile for the 1st quadrant}
\tablehead{   
\colhead{Parameter} & \colhead{2 Jy} & \colhead{3 Jy} & \colhead{4 Jy}\\
}
\startdata
$R_{pk1}$ & 5.17  (kpc) & 5.17 (kpc) & 5.16 (kpc) \\
$R_{pk2}$ & 3.70  (kpc) & 3.71 (kpc) & 3.69  (kpc) \\
$\sigma_{r1}$ & 0.83 (kpc) & 0.82 (kpc)  & 0.82 (kpc) \\
$\sigma_{r2}$ & 0.39 kpc & 0.39 (kpc) & 0.41 (kpc) \\
A & 6.50  & 6.51 & 6.68 \\
\enddata
\end{deluxetable}

\section{The LF in the fourth and first quadrant}

The LFs for the CH87 and Lockman's samples derived by using eq.~(9) and the radial profile
$n(R)$ determined in the previous section are shown in Fig.~5 and 6. The bin size is $\Delta \log L = 0.2$. This choice makes it possible to
perform direct comparisons with LFs obtained, with the same binning, by other
authors (e.g. Smith $\&$ Kennicutt 1989 or KEH89).

\begin{figure*}
\includegraphics[width=4.5in, height=6.5in, ,angle=90]{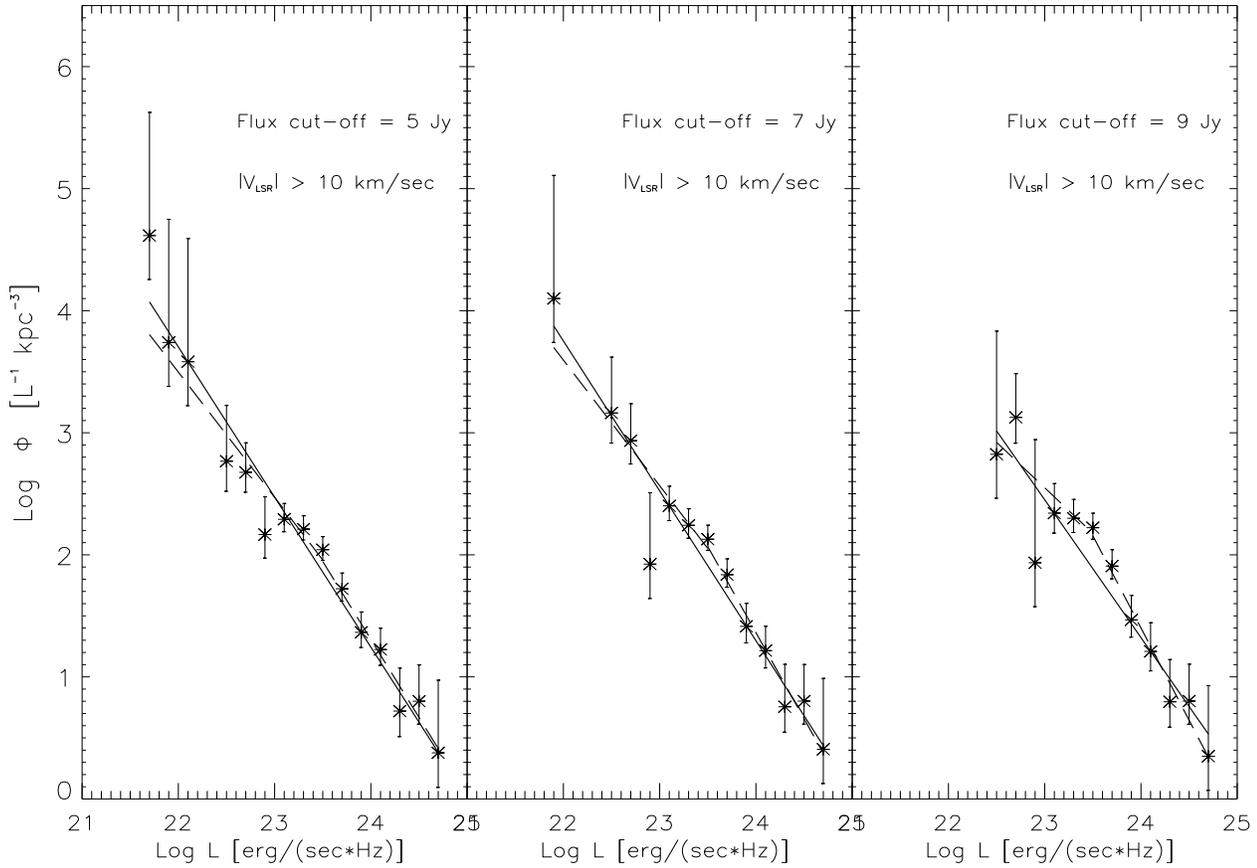}
\caption{LFs in the 4th quadrant for different completeness levels. The solid line indicates the best-fit LF. 
Also shown (dashed line) is a two-component power-law fit. 
}
\end{figure*}

The errors in each bin are computed by applying the formulae by Gehrels (1986) (see \S~3). In this case, 
the effective number of sources per bin, $n_{\rm eff}$, is given by: 
\begin{equation}
n_{\rm eff} = \frac{\left({\sum_{i} {\frac{1}{V{_{\rm eff}}_{i}}}}\right)^2}{\sum_{i}\left({{\frac{1}{V{_{\rm eff}}_{i}}}}\right)^2}
\end{equation}

\begin{figure*}
\includegraphics[width=4.5in, height=6.5in, ,angle=90]{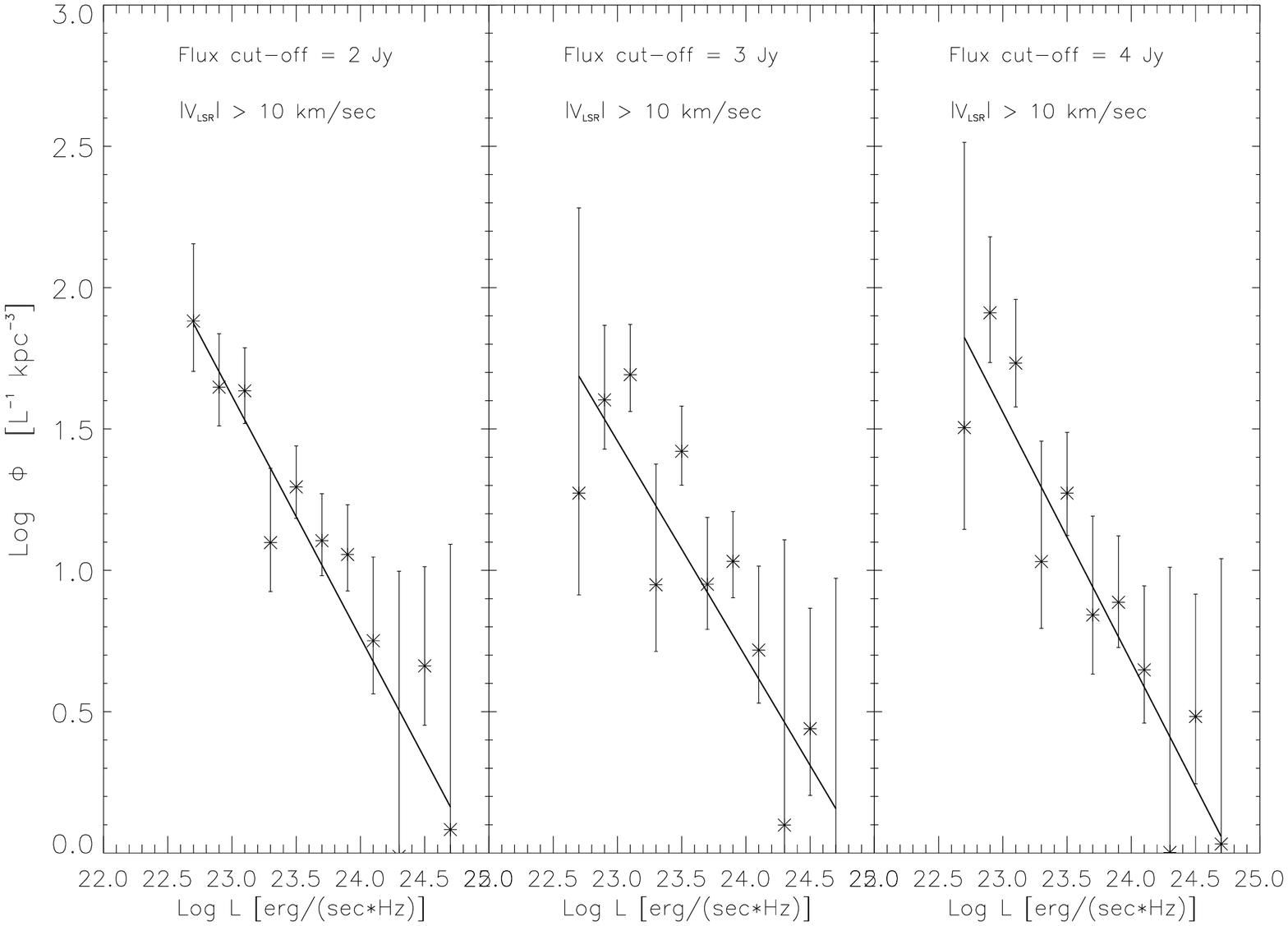}   
\caption{LFs in the 1st quadrant for different completeness levels. The solid line indicates the best-fit LF.}                    
\end{figure*}

\begin{deluxetable}{c|ccc}   
\tabletypesize{\footnotesize}
\tablewidth{0pt}
\tablecaption{Best-fit values for the LFs in the 4th quadrant for different levels
of completeness (single power-law).}
\tablehead{
\colhead{Parameter} & \colhead{5 Jy} & \colhead{7 Jy} & \colhead{9 Jy}\\
}
\startdata
$\alpha$ & 2.23$\pm$0.07 & 2.22$\pm$0.09 & 2.13$\pm$0.12 \\
$\chi^{2}$ & 19.49 & 17.34  & 27.81 \\
$\chi_{r}^{2}$ & 1.49 & 1.58  & 2.78\\
\enddata
\end{deluxetable}

\begin{deluxetable}{c|ccc}
\tabletypesize{\footnotesize}
\tablewidth{0pt} 
\tablecaption{Best-fit values for the LFs in the 1st quadrant for different levels
of completeness (single power-law).}
\tablehead{
\colhead{Parameter} & \colhead{2 Jy} & \colhead{3 Jy} & \colhead{4 Jy}\\
}
\startdata
$\alpha$ & 1.85$\pm$0.11 & 1.76$\pm$0.14 & 1.88$\pm$0.12 \\
$\chi^{2}$ & 11.97 & 23.02  & 11.53\\
$\chi_{r}^{2}$ & 1.33 & 2.56  & 1.28\\
\enddata
\end{deluxetable}

A single power-law fit to the data retrieves the slopes provided, for each quadrant and flux cut-off, 
in Tables 4 and 5. Quoted errors are retrieved from the fitting procedure and are only indicative of the actual errors. 
We have also tried to fit the data with a 
two-component power-law of the form: 

\begin{equation}\label{eq:LF}
\phi(L) = A\left\{\begin{array}{ll}
\left(L/L_{\rm knee}\right)^{-\alpha^{\prime}+1} & \mbox{if $L <L_{\rm knee}$} \\
\left(L/L_{\rm knee}\right)^{-\beta^{\prime}+1} & \mbox{if $L > L_{\rm knee}$} \end{array} \right.\ .
\end{equation}
%


\begin{deluxetable}{c|ccc}
\tabletypesize{\footnotesize}
\tablewidth{0pt}
\tablecaption{Best-fit values for the LFs in the 4th quadrant for different levels
of completeness (two-component power-law).}
\tablehead{
\colhead{Parameter} & \colhead{5 Jy} & \colhead{7 Jy} & \colhead{9 Jy}\\
}
\startdata
$\alpha^{\prime}$ &  2.02$\pm$0.11 & 2.03$\pm$0.15 & 1.74$\pm$0.22 \\
$\beta^{\prime}$ & 2.27$\pm$0.19 & 2.41$\pm$0.20 & 2.51$\pm$0.20 \\
$\log L_{\rm knee}$ & 23.45$\pm$0.11 & 23.54$\pm$0.08 & 23.47$\pm$0.12\\
$\log A$ & 2.13 & 2.19 & 2.04 \\
$\chi^{2}$ & 13.43 & 9.88  & 8.69\\
$\chi_{r}^{2}$ & 1.22 & 1.09  & 1.08\\
\enddata
\end{deluxetable}

From Table 6 and 7, we see that, for the CH87 sample, since $\Delta \chi^{2} \ge 2.3$ (see, e.g., Bevington $\&$ Robinson 1992), 
the addition of two parameters in the fit is justified and the two-component power-law model appears to be favored 
for all completeness levels, while for the Lockman (1989) sample, the single power-law model always 
corresponds  
to the best $\chi^{2}$ value. 

\begin{deluxetable}{c|ccc}
\tabletypesize{\footnotesize}
\tablewidth{0pt}
\tablecaption{Best-fit values for the LFs in the 1st quadrant for different levels
of completeness (two-component power-law).}
\tablehead{
\colhead{Parameter} & \colhead{2 Jy} & \colhead{3 Jy} & \colhead{4 Jy}\\
}
\startdata
$\alpha^{\prime}$ &  1.71$\pm$0.12 & 1.68$\pm$0.14 & 1.96$\pm$0.15 \\
$\beta^{\prime}$ & 1.98$\pm$0.33 & 2.22$\pm$0.35 & 1.93$\pm$0.37 \\
$\log L_{\rm knee}$ & 24.05$\pm$0.23 & 23.94$\pm$0.45 & 23.11$\pm$0.53\\
$\log A$ & 1.75 & 1.84 & 1.88 \\
$\chi^{2}$ & 15.47 & 21.47  & 14.24\\
$\chi_{r}^{2}$ & 2.21 & 3.07  & 2.03\\
\enddata
\end{deluxetable}

To facilitate comparisons with published LFs, we convert our monochromatic LF
(in units of $\hbox{erg}\,\hbox{s}^{-1}\,\hbox{Hz}^{-1}$) into Lyman continuum
fluxes, $Q$, (in $\hbox{photons}\,\hbox{s}^{-1}$) and into H$\alpha$
luminosities, L(H$\alpha$), (in $\hbox{erg}\,\hbox{s}^{-1}$). To convert into
Lyman continuum fluxes, we make use of eq.~(6).  

To express the LF in terms of $H\alpha$ luminosities, we use the relation (Osterbrock 1974):
\begin{equation}
{L(H\beta)/h\nu_{H\beta}\over Q} \sim {{{\alpha_{H\beta}^{\rm eff}} (T_{e})}\over{{\alpha_{\beta}}(T_{e})}}
\end{equation}
which implies that the number of photons emitted by the nebula in H$\beta$ is
directly proportional to its Lyman continuum flux. The hydrogen recombination
coefficients $\alpha_{H\beta}^{\rm eff}$ and $\alpha_{\beta}$ are tabulated,
for various electron densities and temperatures, by Osterbrock (1974) and
Storey $\&$ Hummer (1995). By assuming an electron density $n_{e} =
10^{2}\,\hbox{cm}^{-3}$, we derive the recombination coefficients at $T_{e} =
6200\,$K (CH87) and $T_{e} = 7400\,$K (Lockman 1989) 
by interpolation over the tabulated values. We adopt the value of 4.08
eV for the energy of a photon emitted in a transition between quantum states $n
=4$ and $n=2$. The H$\alpha$ luminosities are then obtained from $L(H\beta)$ by
interpolating at 6200 K and 7400 K the Balmer decrements compiled by Osterbrock (1974) for
different values of  $T_{e}$ (Figure~7 and 8). 

\begin{figure}[h]
\centering
\plotone{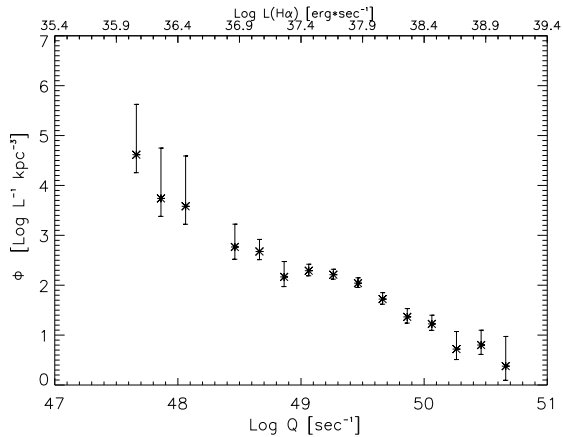}
\epsscale{.6}
\caption{LF in the 4th quadrant expressed in units of Lyman 
continuum flux (lower scale) and $H{\alpha}$ luminosity (upper scale). 
Results refer to a 5 Jy flux cut-off.} 
\end{figure}

\begin{figure}[h]
\centering
\plotone{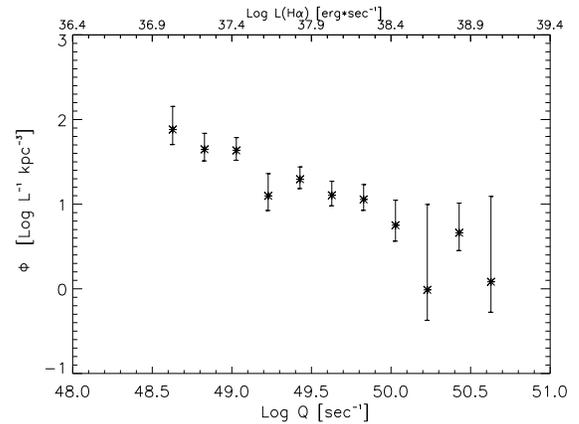}
\epsscale{.6}
\caption{LF in the 1st quadrant expressed in units of Lyman
continuum flux (lower scale) and $H_{\alpha}$ luminosity (upper scale). 
Results refer to a 2 Jy flux cut-off.}
\end{figure}

\subsection{Effect of the choice of the radial density profile}

In this subsection, we discuss how the LF determined through application of
eq.~(9) depends on the adopted radial density profile. To illustrate the effect
of $n(R)$, we compare the LFs  obtained for the CH87 sample by using three
different functional forms, namely: the radial profile of eq.~(12) with the
best-fit values for the parameters in Table~2, an exponential profile, a
constant profile. The reference case for a flux cut-off at 5 Jy is considered. 
We chose the exponential radial profile derived by Kent, Dame
$\&$ Fazio (1991) from the analysis of the Infrared Telescope (IRT) 2.4$\mu$m
data:
\begin{equation}\label{eq:exp}
n(R) = \exp(-R / R_{d})
\end{equation}
with $R_{d}$ = 3 kpc. The expression above is used here only for illustrative
purposes, therefore a discussion regarding the exact value of the radial scale
length, $R_{d}$, is beyond the scope of the present paper.  In the case of a
constant profile, we use the mean of the values tabulated by Bronfman et al.
(2000) for the surface density of OB stars in annuli of increasing
Galactocentric radius (see Table~2 of their paper):
\begin{equation}\label{eq:unif}
n(R) = 1\,\hbox{source}\, \hspace*{0.2truecm} \hbox{kpc}^{-2}\ .
\end{equation}
Once again, the value of the constant is only indicative. Fig.~9 shows the LFs
resulting from using, alternatively, a double-gaussian [eq.~(12)], an 
exponential [eq.~(17)], and a uniform [eq.~(18)] radial
profile. Table~8 provides additional information, i.e. the best-fit values of
the parameters (power-law index and break luminosity) for each curve. We have used a two-component 
power-law model given that, in the previous section, we have shown that it better represents the data 
for the CH87 sample. 

Clearly, the choice of the density profile affects the estimate of 
the luminosity function, although the derived values of the parameters are consistent 
with each other, within the errors.


The different normalizations for the 3 cases reflect the different effective
volumes associated to each density profile. This is also clearly illustrated by
eq.~(7), where one can see that the effective volume acts as a normalization
constant, while the shape of the LF is determined by the observed quantity
$N(L)\,{d\log L}$.

\begin{deluxetable}{lccc}
\tabletypesize{\footnotesize} 
\tablewidth{0pt} 
\tablecaption{Best-fit values
parameters of the LFs for the 4th quadrant (5 Jy flux cut-off) obtained with different radial
density profiles \label{tab:profile}} 

\tablehead{ \colhead{n(R)} &
\colhead{$\alpha' (L< L_{\rm knee})$} & \colhead{$\beta'(L> L_{\rm knee})$} &
\colhead{$\log L_{\rm knee}$}\\
}
\startdata
eq. (12) &  2.19$\pm$0.20 & 2.46$\pm$0.0.15 & 23.45$\pm$0.11  \\
exponential & 1.99$\pm$0.19 & 2.06$\pm$0.42 & 23.47$\pm$0.09 \\
constant & 1.92$\pm$0.12 & 2.98$\pm$0.63 & 23.33$\pm$0.15 \\
\enddata
\tablecomments{Fit for a two component power-law LF [eq.~(\protect\ref{eq:LF})].}
\end{deluxetable}

\begin{figure}[h]
\centering
\plotone{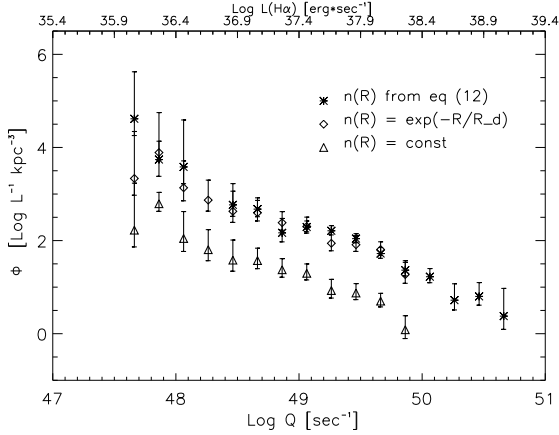}
\epsscale{.6}   
\caption{LF for the CH87 sample (5 Jy flux cut-off) obtained by adopting: a 
double-gaussian radial profile (eq.~(12), stars), 
an exponential profile (diamonds), a constant profile (triangles).} 
\end{figure}

\subsection{Effect of the choice of the rotation curve}

In principle, the LF derived in \S~8 could be dependent on the choice of the
rotation curve model applied to the data to convert the measured line
velocities into Galactocentric and solar distances. As discussed in \S~5, our
adopted rotation curve model is the one by FBS89. To illustrate the possible
dependence of our results on this choice, we recomputed the LF for the CH87
sample by using two alternative models, i.e. the rotation curve by Clemens
(1985) and the one by Brand $\&$ Blitz (1993). The comparison between the LFs
obtained by adopting these models with the one derived by using FBS89 is shown
in Fig.~10. Clearly, the three LFs almost overlap with each other, showing that
our results are not sensitive to different choices for the rotation curve.

\begin{figure}[h]
\centering
\plotone{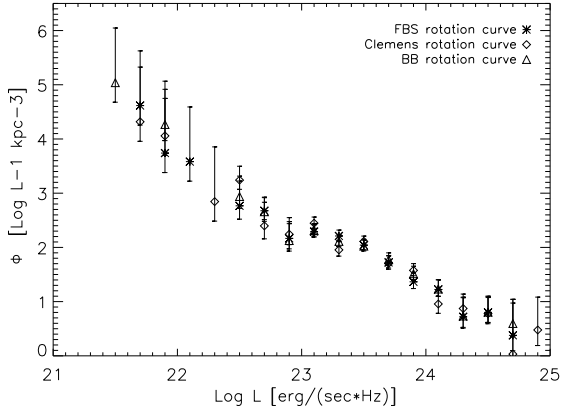}
\epsscale{.6}
\caption{Comparison between LF obtained for the CH87 sample by adopting different
rotation curves. A flux cut-off of 5 Jy is adopted. 
}
\end{figure}

\begin{deluxetable}{c|ccc}
\tabletypesize{\footnotesize}
\tablewidth{0pt}
\tablecaption{Best-fit parameters of the radial profile for the 4th quadrant. 
All sources (including those for which $|V_{LSR}| <$ 10km/s) are considered.}
\tablehead{
\colhead{Parameter} & \colhead{5 Jy} & \colhead{7 Jy} & \colhead{9 Jy}\\
}
\startdata
$R_{pk}$ & 3.21  (kpc) &  2.82(kpc) & 3.13 (kpc) \\
$R_{int}$ & 1.94  (kpc) & 2.08 (kpc) & 1.99  (kpc) \\
$\sigma_{r}$ & 2.85 (kpc)& 3.24 (kpc)  & 3.09 (kpc) \\
\enddata 
\end{deluxetable}

\subsection{Effect of the inclusion of sources with $|V_{LSR}| <$ 10 km/s}

\begin{figure*}
\includegraphics[width=4.5in, height=6.5in, ,angle=90]{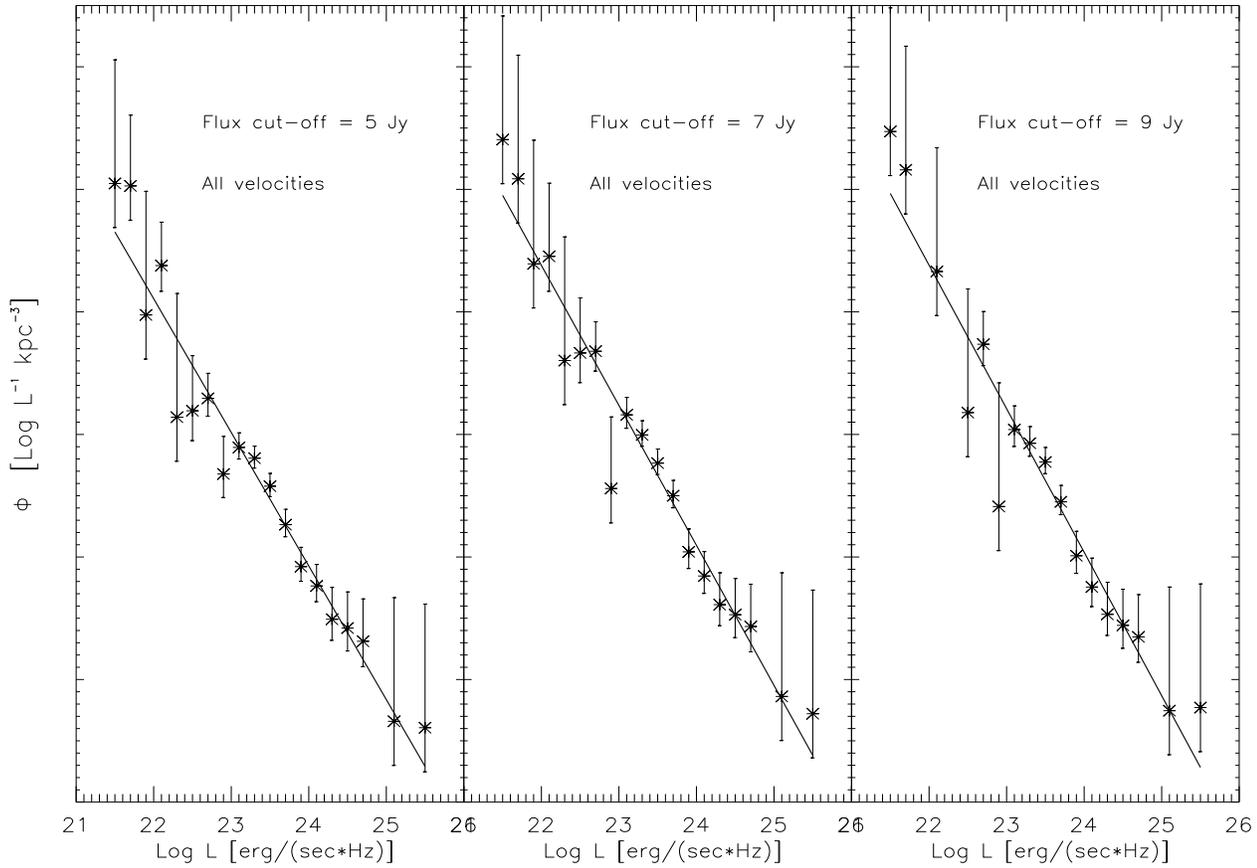}
\caption{LFs in the 4th quadrant for different completeness levels. The solid line indicates the best-fit LF. All sources 
(including those with measured $|V_{LSR}| <$ 10km/s) are considered.}
\end{figure*}

In this section, we investigate how the removal of sources with $|V_{LSR}| <$ 10 km/s affects the derived LFs. 
At this purpose, we recompute, for each sample and considered completeness level, the effective volume, after examination of the 
radial distribution of the sources. These 
are illustrated in Fig.~3 and 4. As noted in \S~5, the removal of the low-velocities sources in the 
fourth quadrant causes a dip 
in the distribution for $R \sim$ 8 kpc. For this reason, instead of applying equation~(12) to fit for $n(R)$, we 
use a gaussian profile with inner cut-off, of the form: 

\begin{eqnarray}\label{eq:1Gaus}
n_{4th}(R) = \exp\left[-((R-R_{pk})/\sigma_{r})^2\right] \times \exp\left[-((R-R_{int})/R)\right]
\end{eqnarray}

For the Lockman (1989) sample, as in \S~7, we adopt eq.~(12), since the addition of the sources with $|V_{LSR}| <$ 10 km/s leaves 
the radial distribution basically unchanged. For illustrative purpose, we provide the best-fit parameters for the radial profile 
(Table~9), as well as 
the slopes (Table~10) of the LFs (Fig.~10), for the fourth quadrant only, given that the CH87 sample is also the on more impacted 
by the presence of sources with low measured velocities. From inspection of Table~10 and Fig.~10 and 
comparison with 
Table~4 and Fig.~5, we see that the slopes are consistent with the values retrieved by removing from the sample  
the low-velocity sources. In the case considered here, though, we find no evidence for a luminosity knee. However, 
such a result should be interpreted with caution, due to the large uncertainty associated to the sources with 
$|V_{LSR}| <$ 10 km/s.

\begin{deluxetable}{c|ccc}
\tabletypesize{\footnotesize}
\tablewidth{0pt}
\tablecaption{Best-fit values for the LFs in the 4th quadrant for different levels
of completeness (single power-law). Sources with $|V_{LSR}| <$ 10km/s are also included.}
\tablehead{
\colhead{Parameter} & \colhead{5 Jy} & \colhead{7 Jy} & \colhead{9 Jy}\\
}
\startdata
$\alpha$ & 2.09$\pm$0.06 & 2.14$\pm$0.06 & 2.17$\pm$0.08 \\
$\chi^{2}$ & 26.96 & 20.55  & 22.40\\
$\chi_{r}^{2}$ & 1.58 & 1.21  & 1.49\\
\enddata
\end{deluxetable}

\section{Discussion}\label{sect:disc}

In the following, we will refer, unless stated otherwise, to the 
reference cases of a flux cut-off at 5 Jy for the CH87 sample,  and at 2 Jy 
for the Lockman (1989) sample. In addition, we consider the case for which sources 
with $|V_{LSR}| <$ 10km/s have been removed from the samples. 

In the fourth quadrant, the H$\alpha$ luminosities of the CH87 sample span $\sim$ 3 orders of magnitude,
from $10^{36}\,$ to $10^{39}\,$ erg/sec or, in equivalent Lyman continuum
luminosities, from $10^{47.6}$ to $10^{50.6}\,\hbox{photons}\,\hbox{s}^{-1}$. The
lower limit corresponds to HII regions ionized by a single star, while the
upper value is indicative of the lack, in our Galaxy, of {\em supergiant} HII regions such as 30 Doradus in the Large Magellanic Clouds,
which originate from very large OB associations.

The LF derived from the CH87 sample presents marginal evidence for a change of
slope at $L_{\rm knee} \sim 
10^{23.45}\,\hbox{erg}\,\hbox{s}^{-1}\,\hbox{Hz}^{-1}$, corresponding to an
H$\alpha$ luminosity of $\sim 10^{37.75}\,$ erg/sec and to a Lyman continuum
luminosity of $\sim 10^{49.41}\,\hbox{photons}\,\hbox{s}^{-1}$ (Fig.~7, upper and lower 
scales). A break in 
the LF of HII regions has been
observed by many authors in external galaxies. Among these, KEH were the first
to conduct a systematic study of the HII region LF in 30 spiral and irregular
galaxies. For 6 galaxies (of Hubble type Sab-Sb) of their sample, they find
that the LF has an abrupt turnover at log $L(H\alpha) = 38.7$--39.0. Recently,
Bradley et al. (2006) have reported a mean value for the break luminosity of
log $L(H\alpha) = 38.6 \pm 0.1$, based on a sample of 56 spiral galaxies. Such
a break has been interpreted in several ways. KEH suggest that the change in
the slope corresponds to a transition between normal HII regions (ionized by a
single star or by a small association) and the class of supergiants. Oey $\&$
Clarke (1997) seem to favor the hypothesis that the break might be caused by
evolutionary effects and a maximum number of stars per cluster, while Beckman
et al. (2000) argue that it is due to  the fact that, at luminosities above a
given threshold, density bounding prevails on ionization bounding.

As for the slope of the LF, our findings are in agreement with published
estimates from extragalactic studies as well as with results obtained for our
own Galaxy. For instance, KEH report, for Sbc-Sc galaxies, a power-law index in
the range 1.5 -- 2.5{\footnote{Retrieved from single power-law fits.}, while
McKee $\&$ Williams and Smith $\&$ Kennicutt (1989) find, respectively, $\alpha
= 1.99 \pm 0.25$ and $\alpha = 2.3 \pm 0.5$ by fitting Lyman continuum
luminosities greater than $10^{49.5}\,\hbox{photons}\,\hbox{s}^{-1}$ and using
the SMB sample. These values are both consistent with the slope obtained for a
single power-law fit, $\alpha = 2.23 \pm 0.07$, as well as with  $\beta' =
2.27\pm0.19$ that we quote in \S\,7 for $L > L_{\rm knee}$, in the case of a double
power-law.

In the first quadrant, the H$\alpha$ luminosities (or
the equivalent Lyman continuum luminosities) of the Lockman et al. (1989) span a smaller range 
of values than the CH87 data set, i.e. from $10^{37}\,$ to $10^{39}\,$ erg/sec or, in equivalent Lyman continuum
luminosities, from $10^{48.6}$ to $10^{50.6}\,\hbox{photons}\,\hbox{s}^{-1}$. In this case, the LF 
is significantly flatter than in the fourth quadrant, suggesting a possible intrinsic difference 
between the HII regions population in the two quadrants, as also hinted by the discrepancy in the 
radial 
profiles (see \S~4.).

The LF obtained for the fourth and first quadrants were used to
estimate the total ionizing luminosity of Galactic HII regions, given by:
\begin{equation}
L_{\rm tot} = V{_{\rm GP}} \int_{L_{\rm min}}^{L_{\rm max}}{n(L) L\, dL}
\end{equation}
where $n(L)= \phi(L)/(L\,\ln 10)$ and $V_{\rm GP}$ is the volume of the Galactic Plane occupied by classical HII regions, i.e.:
\begin{equation}
V_{\rm GP} = 4\pi \int_{0}^{R_{\rm max}}{n(R) R dR}\int_{0}^{\infty}{\rho (z) dz}
\end{equation}
%


\begin{deluxetable}{c|ccc}
\tabletypesize{\footnotesize}
\tablewidth{0pt}
\tablecaption{Total luminosities derived from the LFs in the 4th quadrant}
\tablehead{
\colhead{} & \colhead{5 Jy} & \colhead{7 Jy} & \colhead{9 Jy}\\
}
\startdata
$V_{GP}$ (kpc$^{3}$) & 8.98 & 8.85 & 8.99  \\
$L_{Gal, 5GHz}$  (erg/sec Hz) & 0.95$\times$ 10$^{27}$ & 1.30$\times$ 10$^{27}$ &  0.63$\times$ 10$^{27}$  \\
$L_{Gal, Q}$ ($\hbox{sec}^{-1}$) & 0.87$\times$ 10$^{53}$ & 1.19$\times$ 10$^{53}$ & 0.58$\times$ 10$^{53}$ \\
$L_{Gal, H_{\alpha}}$ (erg/sec)  & 0.19$\times$ 10$^{42}$ & 0.26$\times$ 10$^{42}$ & 0.13$\times$ 10$^{42}$ \\
\enddata
\end{deluxetable}

\begin{deluxetable}{c|ccc}
\tabletypesize{\footnotesize}
\tablewidth{0pt}
\tablecaption{Total luminosities derived from the LFs in the 1st quadrant}
\tablehead{
\colhead{} & \colhead{2 Jy} & \colhead{3 Jy} & \colhead{4 Jy}\\
}
\startdata
$V_{GP}$ (kpc$^{3}$) & 10.49 & 10.77 & 10.45  \\
$L_{Gal, 5GHz}$  (erg/sec Hz) & 1.11$\times$ 10$^{27}$ & 0.82$\times$ 10$^{27}$ &  1.31$\times$ 10$^{27}$  \\
$L_{Gal, Q}$ ($\hbox{sec}^{-1}$) & 0.94$\times$ 10$^{53}$ & 0.70$\times$ 10$^{53}$ & 1.11$\times$ 10$^{53}$ \\
$L_{Gal, H_{\alpha}}$ (erg/sec) & 0.21$\times$ 10$^{42}$ & 0.15$\times$ 10$^{42}$ & 0.24$\times$ 10$^{42}$ \\
\enddata
\end{deluxetable}

with $n(R)$ given by eq.~(12) and the best-fit parameters in Tables 2 and 3, and $\rho(z)$ as in eq.~(10). In addition, $R_{\rm max} =
15\,$kpc, and, for the fourth quadrant, $L_{\rm min}$ = 10$^{21.8}$ erg/sec Hz, $L_{\rm max}$ = 10$^{25}$
erg/sec Hz, while, for the first quadrant, $L_{\rm min}$ = 10$^{22.5}$ erg/sec Hz, $L_{\rm max}$ = 10$^{24.5}$ (see Fig.~5 and 6).   
The total luminosities obtained by application 
of eq.~(20) are shown in Tables 11 and 12. Also 
provided are the values derived for flux completeness levels above the reference cut-off. We note that our estimates 
are lower limits to the actual total luminosities, given that eq.~(20) is integrated between a limited luminosity range. 
From the values quoted in Table~11 and 12, the average of the fourth and first
quadrant is 0.89 $\pm 0.23 \times 10^{53}\,\hbox{sec}^{-1}$. 

The total HII regions Lyman continuum flux for the reference cut-offs can be compared to the total ionizing 
flux of the 
Galaxy. Bennett et al. (1994) find $Q$ = 3.5$\times$ 10$^{53}\,\hbox{sec}^{-1}$, based on COBE observations of the [NII] 205$\mu$m 
line, 
while McKee $\&$ Williams (1997) quote a value of $Q$ = 2.56$\times$ 10$^{53}\,\hbox{sec}^{-1}$. Recently, Murray $\&$ Rahman (2009), from 
the analysis of WMAP data, have estimated  $Q$ = 3.2$\times$ 10$^{53}\,\hbox{sec}^{-1}$. Remarkably, these numbers account for 
the combined  contributions of individual HII regions and the diffuse warm ionized 
gas (WIM). In particular, HII regions are likely responsible for only a fraction (between 10 and 30\%) of the total emission, as 
found, among other authors, in our analysis of the 5-GHz Galactic Plane free-free emission (Paladini, deZotti, Davies $\&$ Giard, 2005). 
Of the HII regions contribution, a significant amount is supposed to originate in sub-giant, or {\em small} HII regions, as 
noted, for instance, by McKee $\&$ Williams (1997). In particular, giant and sub-giant HII regions
are likely to contribute 20$\%$ of the total ionizing luminosity of the Galaxy, while another $\sim$ 20$\%$ is supposedly due to
small HII regions (see their Table~3). McKee $\&$ Williams (1997) find a total ionizing
luminosity from giant and super-giant HII regions of  $0.57 \times 10^{53}\,\hbox{sec}^{-1}$ which, scaled by a factor accounting for
the difference in the adopted $R_{0}$ value (8.5 in their case), becomes equal to $0.50 \times 10^{53}\,\hbox{sec}^{-1}$. Taking 
into account the contribution from small HII regions, this number needs to be doubled, to roughly $1 \times 10^{53}\,\hbox{sec}^{-1}$, i.e. 
one third of the total Lyman continuum flux of the Galaxy, consistent with our results.

\section{Conclusions}

We have investigated the LF of Galactic HII regions in the fourth and first quadrants, using the 
samples from the CH87 recombination lines survey for the fourth quadrant, and from the Lockman (1989) 
recombination lines survey for the first quadrant. We have
shown that a good determination of the highly inhomogeneous spatial
distribution of HII regions is essential to obtain an accurate estimate of the
LF: commonly used but inaccurate density profiles affect both the shape and the
normalization of the LF. 

The LF in the fourth quadrant is well-represented by a single power-law with
spectral index $\alpha$ = 2.23$\pm$0.07, although we find that $\chi^{2}$
goodness-of-fit analysis slightly favors a two-component power law with a break
luminosity at $\log [L(H\alpha)/\hbox{erg/s}] = 37.75$. 
The LF in the first quadrant appears to be flatter, and the best-fit is obtained 
in this case for a single power-law with slope $\alpha$ = 1.85$\pm$0.11. 

The inferred total contribution of individual HII regions to the global ionizing luminosity of the Galaxy 
is found to be in agreement with the estimates by McKee $\&$ Williams (1997) and  
roughly equal to one third.

\acknowledgments Support for this work was provided by NASA through an award issued by JPL/Caltech. 
GDZ acknowledges support by ASI contracts I/016/07/0 ``COFIS'', Planck LFI Activity of Phase E2 and 
the Spitzer Space Telescope Enhanced Science Program. The comments by an anonymous referee have lead us 
to discover a serious numerical error in the previous version of the paper and 
have helped to improve it in many respects.

\appendix

\noindent
From eq. $R=(R_0^2+D^2-2DR_0\cos l)^{1/2}$ we obtain:
\begin{equation}\label{eq:Dtheta}
D= R_0 \cos(l) \pm (R^2 - R_0^2\sin^2(l))^{1/2}\ .
\end{equation}
If we call $\theta$ the angle between the direction from the Galactic Center (GC) to the Sun and the direction from the GC to the source (i.e. between $R_0$ and $R$), the sine theorem gives:
\begin{equation}
\sin(\theta)= {D\over R}\sin(l)\ .
\end{equation}
Combining these equations we get:
\begin{equation}
\sin(\theta)=  \left[{R_0\over R}\cos(l) \pm \left(1-\left({R_0\over R}\right)^2\sin^2(l)\right)^{1/2}\right]\sin(l)\ .
\end{equation}
The number of sources detected at distance $R$, within $\Delta R$, from the GC can be written as:
\begin{equation}
{\cal N}(R)\,\Delta R = n(R)\,R\,\Delta R \int_{\theta_{\rm min}(R)}^{\theta{\rm max}(R)}d\theta \int_{z_{\rm min}(R,\theta)}^{z_{\rm max}(R,\theta)}dz\,\rho(z,R) \int_{\log L_{\rm min}(R,\theta)}^{\log L_{\rm max}}d\log L\ \phi(L)\ ,
\end{equation}
where $z_{\rm max,min}(R,\theta)= b_{\rm max,min}D(R,\theta)$ and $L_{\rm min}(R,\theta)=4\pi D^2(R,\theta)S_{\rm lim}$, with $D(R,\theta)$ given by:
\begin{equation}\label{eq:Dtheta}
D=(R_0^2+R^2-2RR_0\cos \theta)^{1/2}\ .
\end{equation}
If we have a sample selected in the longitude range $l_{\rm min}\le l \le l_{\rm max}$ and $R\le R_0\sin l_{\rm max}$ (i.e. $R$ is smaller than the minimum distance between the GC and the line at
$l=l_{\rm max}$, assuming $l< 90^\circ$), we have:
\begin{equation}
\sin\theta_{\rm max, min}(R)=\left\{{R_0\over R}\cos l_{\rm min} \pm \left[1 - \left({R_0\over R}\right)^2\sin^2 l_{\rm min} \right]^{1/2}\right\}\sin l_{\rm min}\ ,
\end{equation}
where $\theta_{\rm max}$ obviously corresponds to the $+$ sign. {\it Note that $\theta_{\rm max} > \pi/2$, so that $\theta_{\rm max}=\pi-\arcsin(\sin\theta_{\rm max})$},
while $\theta_{\rm min}=\arcsin(\sin\theta_{\rm min})$.

If $R> R_0\sin l_{\rm max}$, we do not have contributions to ${\cal N}(R)$ from the range $\theta_1 \le \theta \le \theta_2$, given by:
\begin{equation}\label{eq:theta12}
\sin\theta_{2, 1}(R)=\left\{{R_0\over R}\cos l_{\rm max} \pm \left[1 - \left({R_0\over R}\right)^2\sin^2 l_{\rm max} \right]^{1/2}\right\}\sin l_{\rm max}\ .
\end{equation}
In practice, we split the integral over $\theta$ in 2 parts, from $\theta_{\rm min}$ to $\theta_1$ and from $\theta_2$ to $\theta_{\rm max}$,
where $\theta_{2, 1}$ are given by the eq. (A7)
if $R> R_0\sin l_{\rm max}$ or by $\theta_{1}=\theta_{2}=\pi/2-l_{\rm max}$ otherwise.

\end{document}